\documentclass[aps,prm,twocolumn,longbibliography]{revtex4-1}

\usepackage{amsmath,amssymb,amsfonts,amsthm}
\usepackage{graphicx,bm}
\usepackage{hyperref}

\begin{document}

\title{Quantitative evaluation of the piezoelectric response of unpoled
ferroelectric ceramics from elastic and dielectric measurements:
tetragonal BaTiO$_3$}
\date{}
\author{F. Cordero}
\affiliation{CNR-ISM, Istituto di Struttura della Materia, Area della Ricerca di Roma -
Tor Vergata,\\
Via del Fosso del Cavaliere 100, I-00133 Roma, Italy}

\begin{abstract}
A method for evaluating the piezoelectric response of unpoled ferroelectric
ceramics from elastic and dielectric measurements is proposed and tested on
BaTiO$_3$.
The method is based on the observation that the softening in a ferroelectric
phase with respect to the paraelectric phase is of piezoelectric origin. The
angular averages of the piezoelectric softening in unpoled ceramics are
calculated for ferroelectric phases of different symmetries. The expression
of the orientational average with the piezoelectric and dielectric constants
of single crystal tetragonal BaTiO$_3$ from the literature reproduces well
the softening of the Young's modulus of unpoled ceramic BaTiO$_3$, after a
correction for the porosity. The agreement is good in the temperature region
sufficiently far from the Curie temperature and from the transition to the
orthorhombic phase, where the effect of fluctuations should be negligible,
but deviations are found outside this region, and the reason for this is
discussed. This validates the determination of the piezoelectric response by
means of purely elastic measurements on unpoled samples. The method is
indirect and, for quantitative assessments, requires the knowledge
of the dielectric tensor. On the other hand, it does not require poling
of the sample, and therefore is
insensitive to inaccuracies from incomplete poling, and can even be used
with materials that cannot be poled, for example due to excessive electrical
conductivity. While the proposed
example of the Young's modulus of a ceramic provides an orientational
average of all the single crystal piezoelectric constants, a Resonant
Ultrasound Spectroscopy measurement of a single unpoled ceramic sample
through the ferroelectric transition can in principle measure all the
piezoelectric constants, together with the elastic ones.
\end{abstract}

\maketitle


\section{Introduction}

The determination of the piezoelectric constants of ferroelectric materials
can be made by piezoelectrically exciting the resonances of samples with
appropriate shapes or by directly quasistatically measuring the charge of a
stressed sample or its strain after application of an electric field.\cite%
{JCJ71,FB13} In all cases the samples must be poled, and the measured
coefficients depend, among other things, on the degree of polarization,
which may be far from complete (see \textit{e.g.} Refs. 16-22 of Ref. \cite%
{TFC14}). As a consequence, the sets of materials constants obtained from
sets of samples with different geometries are often unreliable and not
self-consistent, especially at high temperature, where the problem of
depoling is more acute.\cite{TC15} In extreme situations poling may be
impossible, for example when searching for new multiferroic materials by
modifying ferroelectric materials with magnetic ions, and further optimization
would be required in order to lower an excessive electrical conductivity.

In what follows it will be shown that purely elastic measurements through
the ferroelectric transition on unpoled samples may provide information on
the intrinsic piezoelectric coupling, particularly useful in cases where
poling is difficult or impossible. It can be shown that, within the Landau
theory of phase transitions, the softening in the ferroelectric (FE) phase
with respect to the paraelectric (PE) phase is of piezoelectric origin, and
can be written in tensorial form as\cite{CCT16}
\begin{equation}
\Delta \mathbf{s}^{\text{piezo}}\mathbf{=d}^{+}\mathbf{:\epsilon }^{-1}%
\mathbf{:d~},  \label{Dspiezo}
\end{equation}%
where $\mathbf{s}$ is the compliance, $\epsilon ^{-1}$ the reciprocal
dielectric permittivity and $\mathbf{d}$ the charge piezoelectric
coefficient linking stress $\mathbf{\sigma }$ and polarization $\mathbf{P}$
as $\mathbf{P=d:\sigma }$ (for the notation see Sect. \ref{sectPS}).
According to this relation, knowing the dielectric tensor, it is possible to
evaluate the piezoelectric response by means of purely elastic measurements,
also on unpoled samples, by subtracting the extrapolation of the compliance
of the PE phase. The elimination of the poling process, would be
particularly beneficial with materials with high coercive fields or leakage
currents, and eliminates the uncertainty on the degree of polarization. The
method can be easily applied to pure FE transitions, while, in the presence
of concurrent causes of elastic anomalies, such as octahedral tilting in
perovskites, it would be necessary to separate the
ferroelectric/piezoelectric contribution from the rest. Examples of purely
FE transitions are found in PZT and BaTiO$_{3}$, while in materials based on
Na$_{1/2}$Bi$_{1/2}$TiO$_{3}$ the transitions are of mixed polar and
antiferrodistortive nature,\cite{JT02} so that the softening is both piezoelectric and
from the tilt instability.\cite{139}

Here it will be discussed what kind of information may be obtained
on the piezoelectric coefficients from measurements of the elastic compliance(s)
through the ferroelectric transition on unpoled ceramics. In the usual case
of the Young's modulus (flexural resonance or Dynamic Mechanical Analyzer),
one measures an orientational average of Eq. (\ref{Dspiezo}) in terms
of the single crystal $d_{ijk}$ and $\epsilon _{ij}$
constants, but with a single Resonant Ultrasound Spectroscopy experiment\cite%
{TC15,ZTT17} it is in principle possible to deduce all the elastic and
piezoelectric coefficients. Yet, there are several factors that influence
both the elastic and piezoelectric responses,not necessarily in exactly the
same manner; for example, porosity, texture, grain size and shape,
fluctuations of the order parameter, so that the correspondence between the
effective elastic softening and the piezoelectric response may be not so
straightforward.

In order to make a quantitative check with the intrinsic (single crystal)
materials constants, it is necessary to test a material, whose piezoelectric
properties in the ceramic state are expected to be very close to those of
the single crystal, and for which reliable single crystal material constants
are available. The material of choice is certainly BaTiO$_{3}$, which can be
obtained as ceramic with high purity and large grains, but has the
complication of additional ferroelectric transitions and important
fluctuation effects. It will be verified that the magnitude of the
piezoelectric softening measured in ceramic BaTiO$_{3}$, when passing from
the PE to the FE phase, can be accounted for by the above expression of $%
\Delta \mathbf{s}^{\text{piezo}}$ with the single crystal material
constants, after orientational averaging and a correction for the general
softening caused by porosity. The agreement is good, at least in a
restricted temperature region where the influence of fluctuations from the
neighboring transitions is minimal.

\section{Experimental}

The measurements of the dynamic Young's modulus on ceramic BaTiO$_{3}$ have
already been presented in Ref. \cite{CLM16}, and will be compared with
similar measurements on denser samples from Ref. \cite{CGD96}. The same
labelling BT1 and BT2 as in Ref. \cite{CLM16} is maintained for the two
samples, which were prepared by conventional mixed-oxide powder methods in
different laboratories; here only the relevant details of the preparation
are mentioned. Sample BT1 was prepared with 1~mol\% excess TiO$_{2}$ in the
starting composition, calcined at 1100~$^{\circ }$C for 2~h and sintered in
air at 1400~$^{\circ }\text{C}$ for 1~h, obtaining a mean grain size of
about 50~$\mu \text{m}$. BT1 was cut as a bar of dimensions $33\times
4.2\times 1.1~$mm$^{3}$. Sample BT2 was prepared from stoichiometric amounts
of BaCO$_{3}$ and TiO$_{2}$ powders, calcined in air for 4~h at 1100~$%
^{\circ }$C, compacted in bars by means of isostatic cold pressing and
sintered for 2~h at 1450~$^{\circ }$C. BT2 was cut as a bar of $46\times
4.5\times 0.5$~mm$^{3}$. In both cases no trace of impurity phases was
revealed by powder X-ray diffraction. The densities of BT1 and BT2, measured
with Archimedes' method, were 90\% and 88\% respectively of the theoretical
value of 6.02 g/cm$^{3}$, but the types of porosity were rather different,
judging from the much longer time required by BT2 for stabilizing the weight
when immersed in water with surfactant.

The dynamic Young's modulus $E=$ $E^{\prime }+iE^{\prime \prime }$or the
compliance $s=s^{\prime }-is^{\prime \prime }=$ $1/E$ was measured by
exciting the free flexural resonances of the bars suspended on two thin
thermocouple wires in correspondence with the nodal lines of the first
flexural mode (at $0.225l$ from the ends of the bar with length $l$). Firm
fixing to the wires and shorting of the thermocouple were achieved by
applying silver paint to a sample's edge. The vibration at frequency $f$ is
electrostatically excited by the application of an alternate voltage with
frequency $f/2$ to the electrode. The same electrode is part of a resonating
circuit whose high frequency ($\sim 12$~MHz) is modulated by the sample
vibration, which can be detected with a frequency modulation technique.\cite%
{CDC09}

For samples whose length $l$, width $w$ and thickness $t$ satisfy $l\gg $ $%
w\gg $ $t$, the frequency of the first flexural mode is\cite{NB72}
\begin{equation}
f_{1}=1.028\frac{t}{l^{2}}\sqrt{\frac{E}{\rho }}~.  \label{fF1}
\end{equation}

An irregular or non ideal shape or inhomogeneities of the sample may
introduce an error when deducing $E$ from the above formula. A check of the
consistency of the formula can be done by exciting the 3$^{\text{rd}}$ and 5$%
^{\text{th}}$ flexural modes with frequencies $f_{3}$ and $f_{5}$ and
comparing the measured ratios with the ideal values $f_{3}/f_{1}=$ 5.404 and
$f_{5}/f_{1}=$ 13.34.\cite{NB72} For BT2, whose long and thin shape is
closer to ideal, the 3$^{\text{rd}}$ and 5$^{\text{th}}$ modes have
frequencies within 3\% and 0.3\% of the expected values. The higher
deviation of the 3$^{\text{rd}}$ mode may be in part due to the fact that
the suspension wires were fixed at the nodes of the 1$^{\text{st}}$ mode,
which practically coincide with those of the 5$^{\text{th}}$, but are far
from those of the 3$^{\text{rd}}$ mode. The shorter and thicker BT1 could
not be excited on the 5$^{\text{th}}$ mode, and $f_{3}/f_{1}$ was 5.11, with
a deviation of nearly 6\%. From this check it can be concluded that the
error on the evaluation of the effective (uncorrected for porosity) $E$ of BT2,
due to imperfections in the sample shape and inhomogeneities is within 3\%,
smaller than the uncertainty from the determination of its size.
Sample BT1 will not be used for quantitative purposes.

\section{Results}

\begin{figure}[tbh]
\includegraphics[width=8.5 cm]{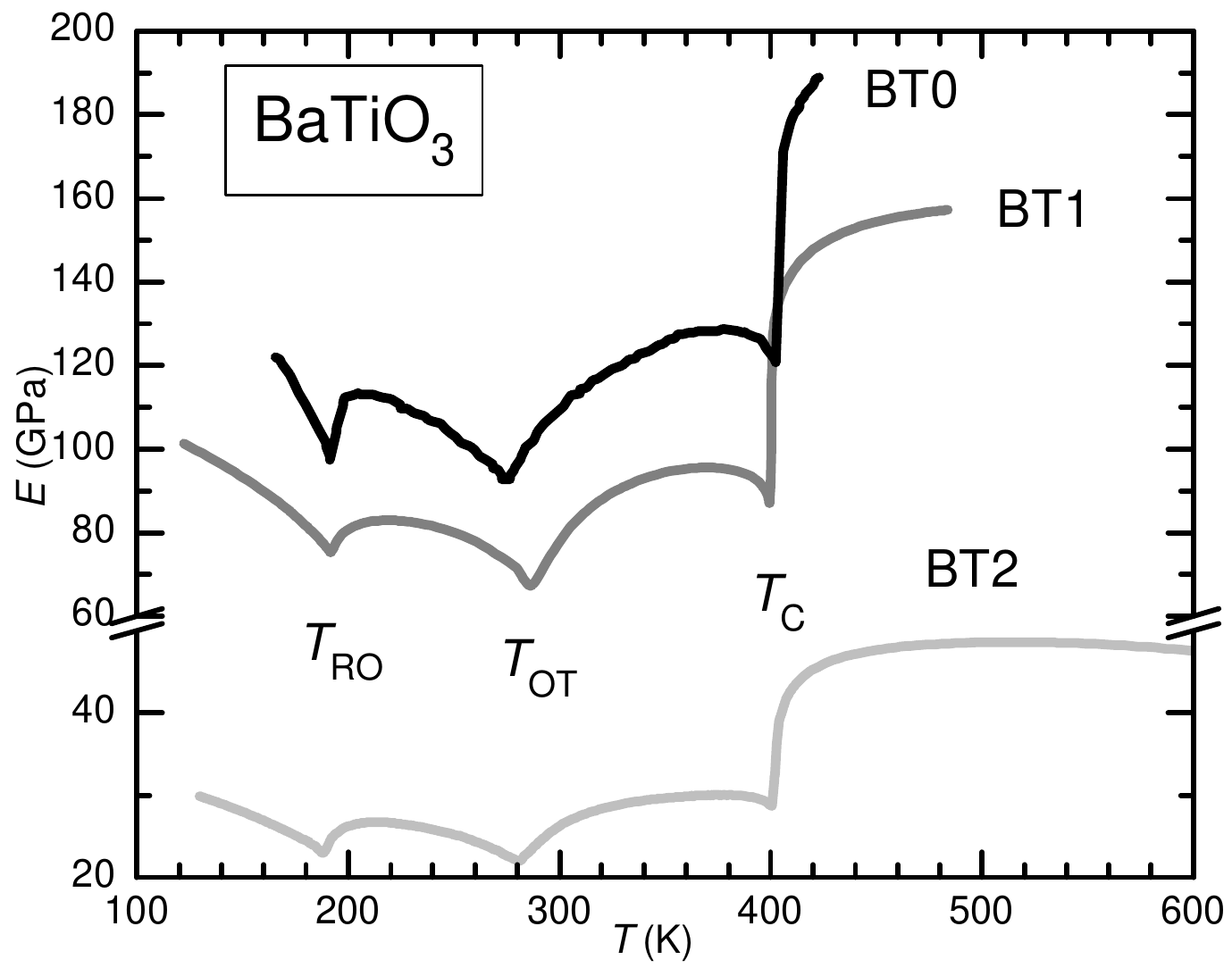}
\caption{Young's modulus of three BaTiO$_{3}$ ceramic samples differently
prepared. Sample BT0 is the coarse grain sample in Fig. 3 of Ref.
\protect\cite{CGD96} (heating 1~K/min), while BT1 and BT2 (cooling $<1.5$%
~K/min in the FE phases) are from Ref. \protect\cite{CLM16}.}
\label{fig_Mall}
\end{figure}

Figure \ref{fig_Mall} shows the Young's moduli of the three samples BT0, BT1
and BT2, measured on the first free flexural mode as a function of
temperature. BT0 is the coarse grain sample of Ref. \cite{CGD96}, and is
used as a reference for the absolute value of the Young's modulus in the PE
phase, because of its highest density. The maximum resonance frequencies
near 500~K in the PE\ phase are 3, 0.9 and 6~kHz respectively and the
moduli are deduced from Eq. (\ref{fF1}), without corrections for the
porosity.

\begin{figure}[tbh]
\includegraphics[width=8.5 cm]{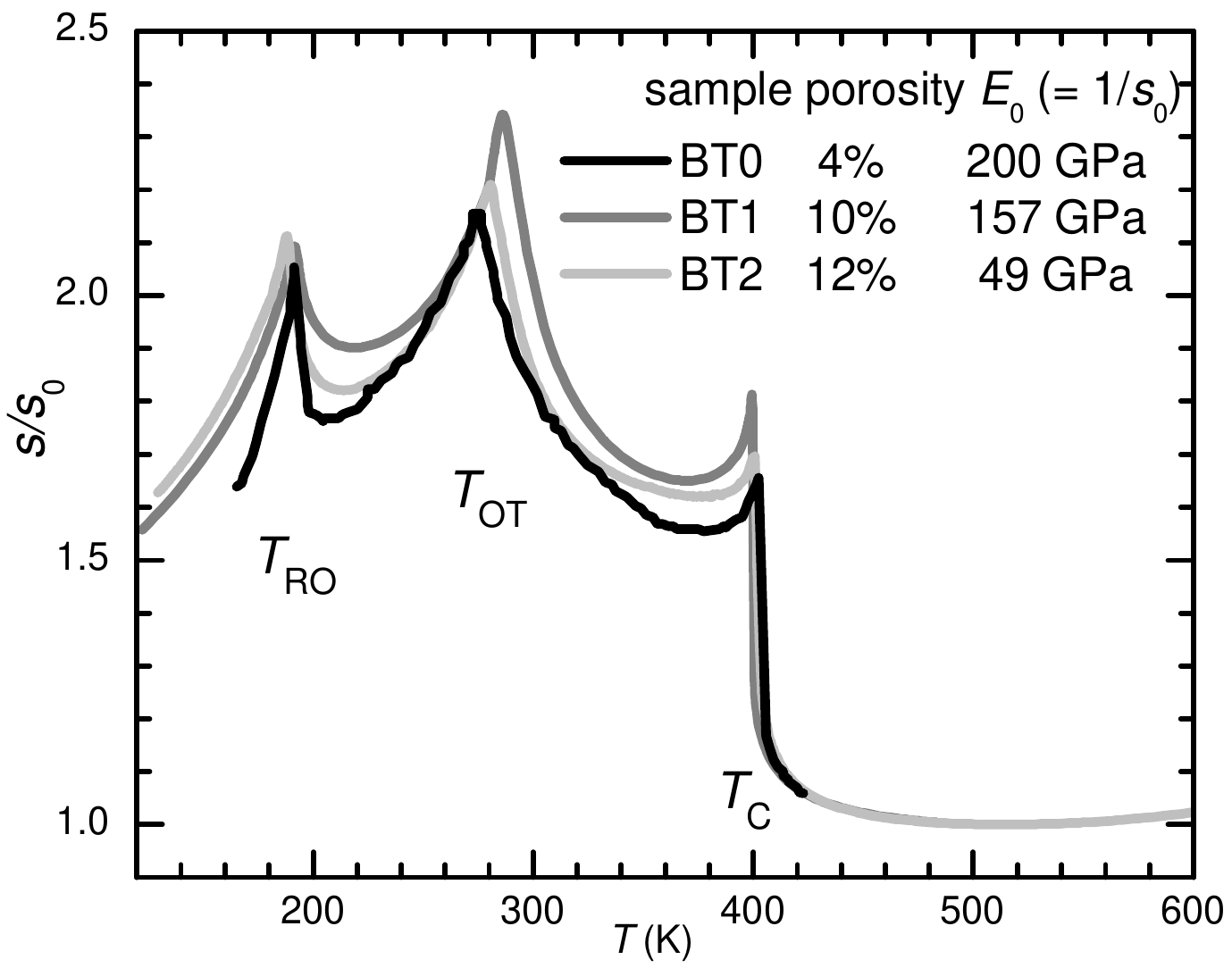}
\caption{Reciprocal Young's modulus $s=E^{-1}$ of the three BaTiO$_{3}$
samples of Fig. \protect\ref{fig_Mall}, normalized with respect to the
(extrapolated) maximum values of $E_{0}$ in the PE phase indicated in the
legend. Also indicated are the porosities evaluated with Archimedes' method.}
\label{fig_sNorm}
\end{figure}

In spite of the widely different values of the Young's moduli, the shapes of
the curves versus temperature are very similar. This is demonstrated in Fig. %
\ref{fig_sNorm}, reporting the compliances $s=E^{-1}$ normalized to $%
s_{0}=1/E_{0}$, where $E_{0}$ is the maximum value in the PE phase; $E_{0}$
is directly taken from the highest value of the $E\left( T\right) $ curve
for BT2 and extrapolated for BT0 and BT1 in order to let the three curves
coincide in the PE phase.

The transition temperature $T_{\mathrm{C}}$ between PE and T-FE phases,
taken at half of the step, is $\sim 400.5$~K ($\sim 1$~K less at the maximum
of $s\left( T\right) $) for BT1 and BT2 and 403~K for BT0, and the
difference is mainly due to the thermal hysteresis, because BT1 and BT2 were
measured during cooling while BT0 during heating. These values of $T_{%
\mathrm{C}}$ are characteristic of stoichiometric and defect free BaTiO$_{3}$%
,\cite{LLK07} and indicate that the differences in the magnitudes of the
elastic moduli are extrinsic, due to differences in the amount and type of
the porosity. As usual, and as explained in terms of the Landau theory,\cite%
{CLM16} the elastic anomalies in the Young's modulus of BaTiO$_{3}$ consist
essentially of a step at $T_{\mathrm{C}}$, between the cubic PE and
tetragonal (T) FE phases, and of a peak at the next transition to the
orthorhombic (O) FE phase at $T_{\mathrm{OT}}$, having the transverse
polarization as new order parameter. There is no simple approximated
description for the final transition to the rhombohedral FE phase at $T_{%
\mathrm{RO}}$, where the components of the order parameter have large
values both in the high and low temperature sides. The marked difference
between the curve BT0 and the curves BT1 and BT2 around 200~K is due to the
hysteresis between heating and cooling (see Fig. 1 in Ref. \cite{CLM16}).

\section{Discussion}

The description of the PE/FE\ transition as a step in the reciprocal Young's
modulus is an oversimplification, since there is a precursor softening
extending for about 200~K above $T_{\mathrm{C}}$, and a sharp upturn of $%
s\left( T\right) $ on approaching $T_{\mathrm{C}}$ from below. In general,
precursor softening is due to the fluctuations of the order parameter, which
are not included in the Landau theory,\cite{SL98,CS98} and compromise its
validity also below the transition, to some extent \cite{SL98}; BaTiO$_{3}$
in particular has quite extended and peculiar fluctuations phenomena in the
PE phase, as briefly discussed in Sect. \ref{sect_fluct}. The upturn below $%
T_{\mathrm{C}}$ may be partly explainable within the Landau theory, by
including the additional terms in the free energy expansion necessary to
describe the other transitions at lower temperature (see \textit{e.g.} Fig.
2(a) in Ref. \cite{II99b}), and, in addition, by the contribution of the
domain walls relaxation. The latter should also exhibit a dispersion in
frequency, which is evident at low frequencies,\cite{CGF96} but hardly
observed in the present measurements at kHz.\cite{CLM16} It is possible
that, analogously to the precursor softening above $T_{\mathrm{C}}$, there
is a precursor softening also above $T_{\mathrm{OT}}$, where the transverse
polarization, acting as order parameter, has mean value zero and may well
fluctuate.

In what follows the steps and approximations necessary for a quantitative
analysis of elastic measurements as those in Figs. \ref{fig_Mall} and \ref%
{fig_sNorm} in terms of single crystal piezoelectric and dielectric
constants are discussed. The aim is to validate the idea of evaluating the
piezoelectric response from the analysis of the elastic softening through
the FE transition, and also to point out the limitations and \textit{caveats}
of the method.\qquad

\subsection{Piezoelectric softening\label{sectPS}}

\bigskip We remind that the origin of the piezoelectric softening is the
electrostrictive term in the free energy\cite{LG77}
\begin{equation}
G_{\sigma ,P}=-\mathbf{\sigma :Q:PP~,}  \label{G_sP}
\end{equation}%
where $\mathbf{Q}$ is the electrostrictive coupling tensor, with components $%
Q_{ijkl}$, $\mathbf{\sigma }$ the stress tensor with components $\sigma _{ij}
$, $\mathbf{PP}$ is a dyad with components $P_{k}P_{l}$, and the colon
denotes the double dot product over a pair of indexes.\cite{SS82} The
electrostrictive term causes a strain $\delta \mathbf{\varepsilon }=-\frac{%
\partial G_{\sigma ,P}}{\partial \mathbf{\sigma }}=$ $\mathbf{Q:PP}$, where,
in the FE phase with spontaneous polarization $\mathbf{P}_{0}>0$, it is $%
\mathbf{P=}$ $\mathbf{P}_{0}+\mathbf{d\cdot \sigma }$, and $\mathbf{%
\varepsilon }_{0}=$ $\mathbf{Q:P}_{0}\mathbf{P}_{0}$ is the spontaneous
strain. On the application of a small stress $\mathbf{\sigma }$, such that $%
\mathbf{d\cdot \sigma \ll P}_{0}$, the part of $\delta \mathbf{\varepsilon }$
exceeding $\mathbf{\varepsilon }_{0}$ contributes to the softening with $%
\Delta \mathbf{s}^{\text{piezo}}=\frac{\delta \mathbf{\varepsilon
-\varepsilon }_{0}}{\mathbf{\sigma }}=$ $2\mathbf{Q:P}_{0}\mathbf{d}$, where
$\mathbf{P}_{0}\approx \mathbf{P}$. The electrostrictive constant can be
eliminated by using $\mathbf{d=}$ $2\mathbf{\epsilon \cdot Q\cdot P}$, where
$\mathbf{\epsilon }$ is the dielectric permittivity, so that the additional
softening can be rewritten\cite{CCT16} as Eq. (\ref{Dspiezo}), repeated here

\begin{equation*}
\Delta \mathbf{s}^{\text{piezo}}\mathbf{=d}^{+}\mathbf{:\epsilon }^{-1}%
\mathbf{:d~.}
\end{equation*}%
Neglecting fluctuations, this term is null in the paraelectric phase, where
at equilibrium $\mathbf{P}=0$, but is directly related to the piezoelectric
constants in the ferroelectric phase. In the simplest case of a second order
FE transition with $\epsilon \propto \left( T_{\mathrm{C}}-T\right) ^{-1}$, $%
P\propto \left( T_{\mathrm{C}}-T\right) ^{1/2}$ and $Q$ independent of $T$, $%
\Delta s^{\text{piezo}}$ is constant in the FE phase, producing a steplike
softening, and $d\propto $ $\left( T_{\mathrm{C}}-T\right) ^{-1/2}$. In such
an ideal case, there is a direct relationship between the amplitude of the
softening below $T_{\mathrm{C}}$ and some combination of the $\mathbf{d}$
and $\mathbf{\epsilon }$ components, depending on the type of the measured
elastic modulus and crystal symmetry.

\subsection{Orientational average of $\Delta \mathbf{s}^{\text{piezo}}$ for
uniaxial stress}

We wish to evaluate the piezoelectric softening, Eq. (\ref{Dspiezo}), to the
reciprocal Young's modulus $\overline{E^{-1}}=\overline{s}$ in an unpoled
ceramic material. We start from the compliance in a single crystal in the
direction $\mathbf{\hat{n}}$, which is the elongation in that direction due
to a unitary uniaxial stress $\mathbf{\sigma =\hat{n}\hat{n}}$, with
components $\sigma _{ij}=n_{i}n_{j},$ and is given by\cite{SS82,Nye57} $%
E^{-1}\left( \mathbf{\hat{n}}\right) =$ $\mathbf{\hat{n}\hat{n}:s:\hat{n}%
\hat{n}}$. Analogously, the contribution of $\Delta \mathbf{s}^{\text{piezo}}
$ to this compliance is
\begin{equation}
\mathbf{\Delta \mathbf{s}^{\text{piezo}}}\left( \mathbf{\hat{n}}\right)
\mathbf{=\hat{n}\hat{n}:\Delta \mathbf{s}^{\text{piezo}}:\hat{n}\hat{n}~},
\label{Dspiezonn}
\end{equation}%
and it must be averaged over the ceramic sample. Since we are dealing with
untextured unpoled ceramics, we can assume random orientations of the grains
and polarization $\mathbf{P}$. This average is simpler than for a poled
ceramic, where the reorientation of the polarization must be assumed along
the permissible crystallographic direction closer to the direction of the
poling field \cite{Bae57,LFL99,LR07}; in addition, the reorientation may be
far from complete, especially when the switching is also ferroelastic, as in
$90%
{{}^\circ}%
$ switching in the T phase (see \textit{e.g.} Ref. \cite{TFC14} and Refs.
16-22 therein). Yet, taking the angular average of Eq. (\ref{Dspiezo}) is an
oversimplification also in the unpoled state, since it is equivalent to the
Reuss average, which assumes uniform stress across the sample, while the
other limit is the assumption of uniform strain, leading to the Voigt
average of the elastic moduli instead of the compliances. Completely
analogous averages can be made with the electrostrictive constants.\cite%
{HFJ89c} Sometimes, the average of the two methods, known as Hill
polycrystalline average, is used, but more refined and complicated methods
have been developed, that take into account in a self-consistent manner the
reciprocal action of the domains or grains on their neighbors (\textit{e.g.}
Ref. \cite{Ber05b}). Also the shape of the domains and grains has a role,
which can be simulated in the more sophisticated treatments. Since the major
source of extrinsic softening in our samples is porosity, it would be
excessive to adopt such methods, based on largely undetermined parameters
like the distribution of domain/grain sizes and shapes. Therefore, we will
simply perform the angular average of Eq. (\ref{Dspiezo}). This choice has
already been found to provide good results for BaTiO$_{3}$.\cite{BJ58}

It is convenient to adopt the Voigt or matrix notation:\cite{Nye57} 
$ij=11,22,33\rightarrow \alpha =1,2,3$, $ij=12,21\rightarrow \alpha =6$, $%
ij=23,32\rightarrow \alpha =4$, $ij=13,31\rightarrow \alpha =5$, with the
additional rule that the components of $\mathbf{\varepsilon }$, $\mathbf{s}$
and $\mathbf{Q}$ have to be multiplied by 2 for each index $\alpha \geq 4$.
In this manner, for the tetragonal $4mm$ symmetry the $\mathbf{d}$ and $%
\epsilon ^{-1}$ matrices are written as\cite{SS82,Nye57}
\begin{equation*}
\epsilon ^{-1}=\left(
\begin{array}{ccc}
\epsilon _{11}^{-1} & 0 & 0 \\
0 & \epsilon _{11}^{-1} & 0 \\
0 & 0 & \epsilon _{33}^{-1}%
\end{array}%
\right)
\end{equation*}%
\begin{equation*}
d\mathbf{=}\left(
\begin{array}{cccccc}
0 & 0 & 0 & 0 & d_{15} & 0 \\
0 & 0 & 0 & d_{15} & 0 & 0 \\
d_{31} & d_{31} & d_{33} & 0 & 0 & 0%
\end{array}%
\right)
\end{equation*}%
and the piezoelectric softening Eq. (\ref{Dspiezo}) becomes
\begin{equation}
\Delta s^{\text{piezo,T}}\mathbf{=}\left(
\begin{array}{cccccc}
\frac{d_{31}^{2}}{\epsilon _{33}} & \frac{d_{31}^{2}}{\epsilon _{33}} &
\frac{d_{31}d_{33}}{\epsilon _{33}} & 0 & 0 & 0 \\
\frac{d_{31}^{2}}{\epsilon _{33}} & \frac{d_{31}^{2}}{\epsilon _{33}} &
\frac{d_{31}d_{33}}{\epsilon _{33}} & 0 & 0 & 0 \\
\frac{d_{31}d_{33}}{\epsilon _{33}} & \frac{d_{31}d_{33}}{\epsilon _{33}} &
\frac{d_{33}^{2}}{\epsilon _{33}} & 0 & 0 & 0 \\
0 & 0 & 0 & \frac{d_{15}^{2}}{\epsilon _{11}} & 0 & 0 \\
0 & 0 & 0 & 0 & \frac{d_{15}^{2}}{\epsilon _{11}} & 0 \\
0 & 0 & 0 & 0 & 0 & 0%
\end{array}%
\right)   \label{t-DspiezoT}
\end{equation}

The magnitude of $\Delta \mathbf{s}^{\text{piezo}}$ along the generic
crystallographic direction $\mathbf{\hat{n}=}$ $\left( \sin \theta \cos \phi
,\sin \theta \sin \phi ,\cos \theta \right) $, where $\theta $ is the angle
between the direction of the uniaxial stress and the tetragonal $c$ axis, is
\begin{equation*}
\Delta s^{\text{piezo,T}}\left( \mathbf{\hat{n}}\right) =\mathbf{\hat{n}\hat{%
n}:}\Delta \mathbf{s}^{\text{piezo,T}}:\mathbf{\hat{n}\hat{n}=~}nn\cdot
\Delta s^{\text{piezo,T}}\cdot nn,
\end{equation*}%
with
\begin{equation*}
nn=\left(
n_{1}^{2},n_{2}^{2},n_{3}^{2},n_{2}n_{3},n_{1}n_{3},n_{1}n_{2}\right) ,
\end{equation*}%
and we obtain
\begin{align*}
\Delta s^{\text{piezo,T}}\left( \mathbf{\hat{n}}\right) & =\frac{%
d_{15}^{2}\left( 1-n_{3}^{2}\right) n_{3}^{2}}{\epsilon _{11}}+ \\
& \frac{d_{31}^{2}\left( 1-n_{3}^{2}\right) ^{2}+2d_{31}d_{33}\left(
1-n_{3}^{2}\right) n_{3}^{2}+d_{33}^{2}n_{3}^{4}}{\epsilon _{33}}~.
\end{align*}%
where $n_{1}^{2}+n_{2}^{2}+n_{3}^{2}=1$ and $\overline{n_{1}^{2}}=\overline{%
n_{2}^{2}}$ have been used. The angular averages are $\overline{n_{3}^{2}}=%
\frac{1}{\pi }\int_{0}^{\pi }d\theta ~\cos ^{2}\theta =\frac{1}{3}$ and $%
\overline{n_{3}^{4}}=\frac{1}{5}$, so that finally
\begin{equation}
\overline{\Delta s^{\text{piezo,T}}}=\frac{%
8d_{31}^{2}+4d_{31}d_{33}+3d_{33}^{2}}{15\epsilon _{33}}+\frac{2d_{15}^{2}}{%
15\epsilon _{11}}  \label{av-DspiezoT}
\end{equation}

In the same manner it is possible to find the expressions valid for the
orthorhombic $mm2$ (O) and rhombohedral $3m$ (R) symmetries\cite{Abr09} of
the other FE phases of BaTiO$_{3}$:

\begin{equation}
\overline{\Delta s^{\text{piezo,R}}} =\frac{%
8d_{31}^{2}+4d_{31}d_{33}+3d_{33}^{2}}{15\epsilon _{33}}+%
\frac{\left( 2d_{15}^{2}+4d_{22}^{2}\right) }{15\epsilon _{11}} \label{av-DspiezoR}
\end{equation}

\begin{equation}
\begin{split}
\overline{\Delta s^{\text{piezo,O}}} =\frac{%
3d_{31}^{2}+3d_{32}^{2}+3d_{33}^{2}}{15\epsilon _{33}}+  \\
\frac{2d_{31}\left( d_{32}+d_{33}\right) +2d_{32}d_{33}}{15\epsilon _{33}} +%
\frac{d_{15}^{2}}{15\epsilon _{11}}+\frac{d_{24}^{2}}{15\epsilon _{22}}
 \label{av-DspiezoO}
\end{split}
\end{equation}

\subsection{Influence of fluctuations\label{sect_fluct}}

The expression (\ref{Dspiezo}) of $\Delta s^{\text{piezo}}$ is valid for
free energy expansions containing as many powers of $\mathbf{P}$ as
necessary for describing sequences of phase transitions, as done for example
in Refs. \cite{LCC05,WTD07}, unless there are other types of coupling
between stress and strain besides the electrostrictive one, Eq. (\ref{G_sP}).
Therefore, within the Landau theory, it should be possible to reproduce also
the upturn of $s$ below $T_{\mathrm{C}}$ and the cusps at $T_{\mathrm{OT}}$
and $T_{\mathrm{RO}}$, as done in the simulations of Ishibashi and Iwata,%
\cite{II99b} and Eq. (\ref{Dspiezo}) would automatically connect the
elastic, dielectric and piezoelectric responses. In particular, it has been
shown\cite{CLM16} that the linearization of the shear electrostrictive term
with respect to the small transverse component of $\mathbf{P}$ causes a cusp
of $s\left( T\right) $ at $T_{\mathrm{OT}}$; that is exactly piezoelectric
softening\textbf{. }However, the Landau theory, and hence also Eq. (\ref%
{Dspiezo}), do not take into account the fluctuations of the polarization,
which certainly play an important role in the PE phase and, presumably, also
above the next transition at $T_{\mathrm{OT}}$.

The fluctuations may renormalize the coefficients of the free energy
expansion. transforming a second order transition into first order,\cite%
{SL98} and this should not affect the validity of Eq. (\ref{Dspiezo}), but
they may also produce different contributions to the dielectric
susceptibility and to the compliance, so that the above expressions of $%
\Delta s^{\text{piezo}}$ are not expected to be valid in some temperature
range near the transition temperatures. Determining a criterion for the
temperature range of validity of the Landau theory in the presence of
fluctuations, or including the effect of fluctuations in the expression (\ref%
{Dspiezo}) is outside the scope of the present paper. General criteria for
estimating the range of validity of the Landau theory are discussed for
example in Ref. \cite{RAT07}. Here we only remind that the longitudinal
fluctuations of the polarization should be partially suppressed by the
accompanying depolarizing fields, which are absent in the transverse
fluctuations.\cite{SL98,RAT07} Therefore, it is expected that different
transitions are differently affected by fluctuations: the FE/PE transition
at $T_{\mathrm{C}}$ should be accompanied by prevailing longitudinal
fluctuations, while that at $T_{\mathrm{OT}}$, consisting in a rotation of
the polarization, should be accompanied by prevalent transverse
fluctuations. The transition at $T_{\mathrm{RO}}$ should not be much
affected by fluctuations, since both the longitudinal and transverse
components of the polarization are finite. In spite of these considerations
(expected quenching of the longitudinal fluctuations above $T_{\mathrm{C}}$ by the
depolarizing fields), important precursor softenings have been measured in
PE BaTiO$_{3}$ with various techniques, such as ultrasound velocity,\cite%
{KHI73} Resonant Ultrasound Spectroscopy\cite{ACS13} and Brillouin
scattering,\cite{KKR11} and Fig. \ref{fig_bkg}(a) confirms that the
precursor softening starts already at $T_{\mathrm{C}}+250$~K. Other
precursor phenomena have been observed in the Raman spectrum,\cite{FL72}
refraction index,\cite{BD82} Second Harmonic Generation,\cite{PKS12} and all
these phenomena are usually discussed in terms of polar nanoregions.\cite%
{KKR11,BD82,PKS12} The piezoelectric coupling is no exception, since the
mechanical resonances can be piezoelectrically excited up to 600~K.\cite%
{ACS13}

Without a precise knowledge of the nature of the fluctuations in BaTiO$_{3}$
and of their effect on Eq. (\ref{Dspiezo}), it seems safe to apply this
formula in a region far from both $T_{\mathrm{C}}$ and $T_{\mathrm{OT}}$;
the best choice seems the plateau of $s\left( T\right),$ with shallow
minimum at $T_{m}=$ 373~K. The verification of Eq. (\ref{Dspiezo}) will be
first made at that temperature.

\subsection{Correction for the porosity}

In order to compare the magnitude of the softening below $T_{\mathrm{C}}$
with Eq. (\ref{Dspiezo}), where the single crystal materials constants are
insterted, the overall softening effect of porosity must be taken into
account. The softening from porosity depends on both the volume fraction of
the pores (or from the relative sample density) and on their shape. The
effect of shape may be very large: passing from spherical to flat lenticular
or "penny shaped" pores, keeping their volume fraction constant, may soften
the sample by several times.\cite{Wu66,Dun95} Therefore, the evaluation of
the softening from the relative density and the visual characterization of
the pore shapes is rather aleatory. Figures \ref{fig_Mall} and \ref%
{fig_sNorm} clearly demonstrate this point, since the relationship between
the porosity, as volume fraction, and the Young's modulus in the PE phase
cannot be expressed with a linear or smooth function. The fact that BT2 with
12\% porosity is more than three times softer than BT1 with 10\% porosity is
certainly due to a morphology of its porosity similar to cracks and
fissures. This is confirmed by the fact that, when measuring the densities
with the Archimedes method, the time for stabilizing the weight in water was
much longer for BT2. The penetration of water in the connected porosity also
fictitiously enhances the relative density, compared to non connected
porosity, so that the pores volume fraction of BT2 might be larger than
12\%.

It has already been shown in Fig. \ref{fig_sNorm} that, after rescaling, the
curves of the compliance of BaTiO$_{3}$ versus temperature of samples with
different porosities coincide to a reasonable degree of accuracy. This is true
even for BT2, four times softer than BT0. The denser sample BT0 measured by
Fantozzi has an extrapolated maximum Young's modulus $s_{0}^{-1}=$ 200~GPa,
which we assume to be close to the intrinsic value that would be measured in
a completely dense sample. For the following comparison with single crystal
data, we choose our denser sample BT1, whose maximum extrapolated modulus is
$s_{0}^{-1}=$ 157~GPa, and use a rescaling factor%
\begin{equation}
f=157/200=0.785~,  \label{f}
\end{equation}%
which makes BT1 to overlap with BT0.

The value of 200~GPa for the intrinsic Young's modulus of PE BaTiO$_{3}$
deduced here is in line with the highest values available in the literature:
197~GPa from ultrasound experiments \cite{LLK90}. It is also close to the
value calculated on the basis of the single crystal compliances at 150~%
${{}^\circ}$C: \cite{BJ58} $s_{11}=8.33\times 10^{-12}~$Pa$^{-1}$,
$s_{12}=-2.68\times 10^{-12}~$Pa$^{-1}$, $s_{44}=9.24\times 10^{-12}~$Pa$^{-1}
$. The orientational average of the reciprocal Young's modulus $\overline{%
E^{-1}\left( \mathbf{\hat{n}}\right) }$ for a cubic polycrystal is
calculated in a manner analogous to that used for obtaining Eq. (\ref%
{av-DspiezoT}), and yields\cite{NB72,SS82} $\overline{E^{-1}}=$ $s_{11}-%
\frac{2}{5}\left( s_{11}-s_{12}-\frac{1}{2}s_{44}\right) $; inserting the
single crystal values,\cite{BJ58} one obtains $\overline{E}=214~$GPa.

It results that porosity is a major source of uncertainty in determining the
intrinsic elastic constants, but the same is true for the direct
measurements of the piezoelectric coefficients, though they are not affected
by porosity exactly in the same manner as the elastic constants.\cite{Dun95}
The different effects of porosity on the effective $s$ and $\epsilon $
imply that, for large porosities, the effective $d$ deduced from Eq. (\ref%
{Dspiezo}) may be different from the effective $d$ measured on the same
material by direct methods (without considering possible issues related to
incomplete poling). Moreover, it cannot be excluded that in samples whose
porosities are particularly high or with peculiar morphologies, also the
elastic compliance $s^{0}\left( T\right) $ and piezoelectric softening $%
\Delta \mathbf{s}^{\text{piezo}}\left( T\right) $ are differently affected,
so that a rescaling factor is not sufficient to account for the effect of
porosity.

Summing up, the compliances in the PE phase of the three samples in Fig. \ref%
{fig_sNorm} differ by up to four times, due to porosity, but exhibit a
satisfactory reproducibility, after rescaling. This encourages in correcting
for porosity both compliance and piezoelectric softening with a common
rescaling factor; yet, it cannot be excluded that large porosities with
particular morphologies impair the validity of this simple type of
correction.

\subsection{Subtraction of the background elastic constant}

In order to extract the piezoelectric softening, the other contributions to
the compliance must be subtracted. This task should be easy for a purely
ferroelectric transition, but not in the presence of concomitant
instabilities of other nature. For example, in Na$_{1/2}$Bi$_{1/2}$TiO$_{3}$
the polar transitions are accompanied by changes in the octahedral tilt
patterns, and, in order to use Eq. (\ref{Dspiezo}), one should somehow
identify and subtract the softening connected with these antiferrodistortive
modes. This may be be a major difficulty in general, but it is possible for
the antiferroelectric/ferroelectric transitions in PZT, thanks to the fact
that the polar and antiferrodistortive modes have different kinetics and can
be observed separately in the elastic measurements (Ref. \cite{CTC14} and
see note 12 of Ref. \cite{CCT16c}; note that in those cases one observes the
loss of piezoelectric softening, namely a stiffening, during cooling from
the FE to the AFE phase). In the absence of additional
instabilities, one can reasonably assume that the softening at the FE
transition is totally due to Eq. (\ref{Dspiezo}), besides fluctuations.
Therefore, in the absence of the FE transition, the background compliance
would vary approximately linearly with $T$, due to the phonon anharmonicity
(see \textit{e.g.} the temperature coefficient of the inverse
compressibility in terms of the Gr\"{u}neisen constant, Eq. (4.58) in Ref.
\cite{BH54}, and Ref. \cite{WTL61}).

\begin{figure}[tbh]
\includegraphics[width=8.5 cm]{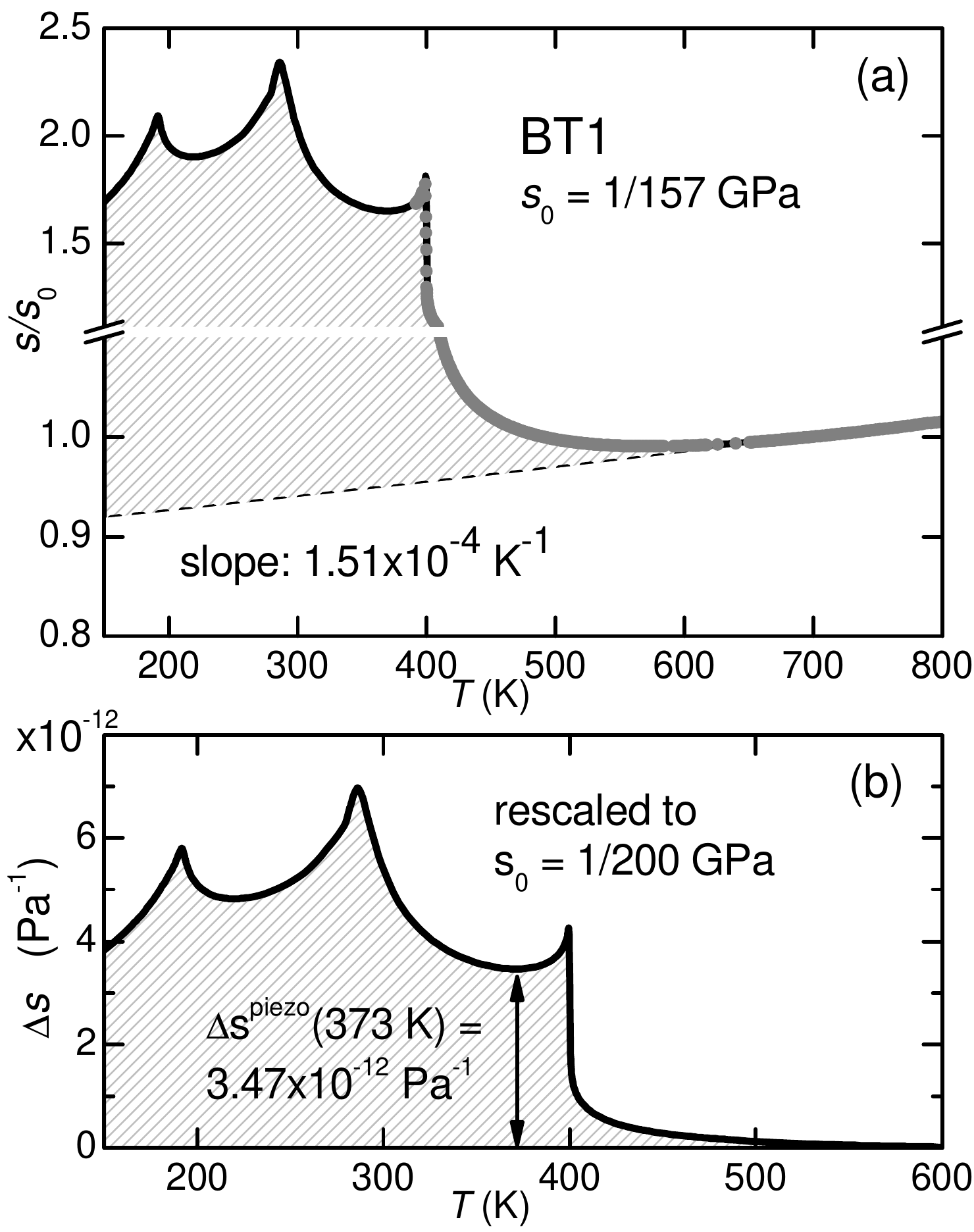}
\caption{(a) Compliance of sample BT1 (continuous line and gray symbols),
normalized with $s_{0}=1/157$~GPa and extrapolation of the linear high
temperature trend (dashed). (b) Absolute value of the compliance exceeding
the linear background after rescaling $s_{0}^{-1}$ from 157 to 200~GPa.}
\label{fig_bkg}
\end{figure}

Figure \ref{fig_bkg} shows two measurements of BT1: the black line up to
480~K is the compliance from which we will deduce the piezoelectric
softening, while the gray symbols above 390~K are another measurement
extended to 800~K. Between 390 and 480~K the two curves overlap perfectly,
and the one at higher temperature can be extrapolated from the linear region
down to below $T_{\mathrm{C}}$. It is necessary to extrapolate the linear
anharmonic behavior of the elastic moduli at such high temperatures, because
the precursor softening extends at least up to 650~K. The resulting dashed
line, with a slope of $1.51\times 10^{-4}$~K$^{-1}$, is subtracted from $%
s\left( T\right) $ in order to obtain $\Delta \mathbf{s}^{\text{piezo}}$.
The resulting compliance, exceeding the compliance of the PE\ phase with the
anharmonic corrections, is rescaled by the factor (\ref{f}), to account for
porosity, and should represent the piezoelectric softening plus
contributions from fluctuations, which are different in $s$, $\epsilon $ and
$d.$ As discussed previously, the temperature region where the effect of
fluctuations should be minimal is the minimum of the plateau around 373~K,
and therefore we choose this temperature for determining the magnitude of
the piezoelectric softening, which results to be
\begin{equation}
\Delta \mathbf{s}_{\text{ceram}}^{\text{piezo}}\left( 373~\text{K}\right)
=3.47\times 10^{-12}~\text{Pa}^{-1}~.  \label{Ds-BT1}
\end{equation}

\subsection{Comparison with single crystal material constants}

In spite of the extensive literature on the piezoelectric and elastic
properties of BaTiO$_{3}$, it is not easy to find a complete set of material
constants, especially as a function of temperature. Schaefer and coworkers%
\cite{SSD86} measured the temperature dependencies of the $s_{\alpha \beta }$%
, $\epsilon _{ij}$ and $d_{i\beta }$ constants within the T phase from a set
of single crystals. Their results should be more reliable than the old ones,%
\cite{BJ58} in view of the higher $T_{\mathrm{C}}$ and presumed reduced
twinning, but some inaccuracies may arise from incomplete poling of some of
the samples.\cite{SSD86} The temperature dependencies of the $s_{\alpha
\beta }$, $\epsilon _{ij}$ and $d_{i\beta }$ constants have also been
calculated \cite{DBS02,BDS03} from a Landau free energy expansion,\cite%
{Bel01} which reproduces the $T-E$ phase diagram and piezoelectric
properties of BaTiO$_{3}$. By inserting the values of those coefficients at
373~K in Eq. (\ref{Dspiezo}) one obtains:
\begin{eqnarray}
\Delta \mathbf{s}_{\text{cryst}}^{\text{piezo}}\left( 373~\text{K}\right)
&=&4.84\times 10^{-12}~\text{Pa}^{-1} \\
&&\text{(Ref. \cite{SSD86})}  \notag \\
\Delta \mathbf{s}_{\text{Landau}}^{\text{piezo}}\left( 373~\text{K}\right)
&=&3.68\times 10^{-12}~\text{Pa}^{-1}~ \\
&&\text{(Refs. \cite{DBS02,BDS03}).}  \notag
\end{eqnarray}%
The value deduced from BT1, Eq. (\ref{Ds-BT1}), is only 6\% lower than that
calculated using the Landau free energy\cite{DBS02,BDS03} and 39\%\ smaller
than using the single crystal data \cite{SSD86}; this can be considered as a
good agreement, in view of the uncertainty from the porosity and sample
shape, and the fact that the reference data differ of 33\% between
themselves. This validates the use of Eqs. (\ref{av-DspiezoT}-\ref%
{av-DspiezoO}) for expressing the magnitude of the softening from the PE to
FE phase in terms of the piezoelectric and dielectric coefficients for the
T, R and O symmetries.

\begin{figure}[tbh]
\includegraphics[width=8.5 cm]{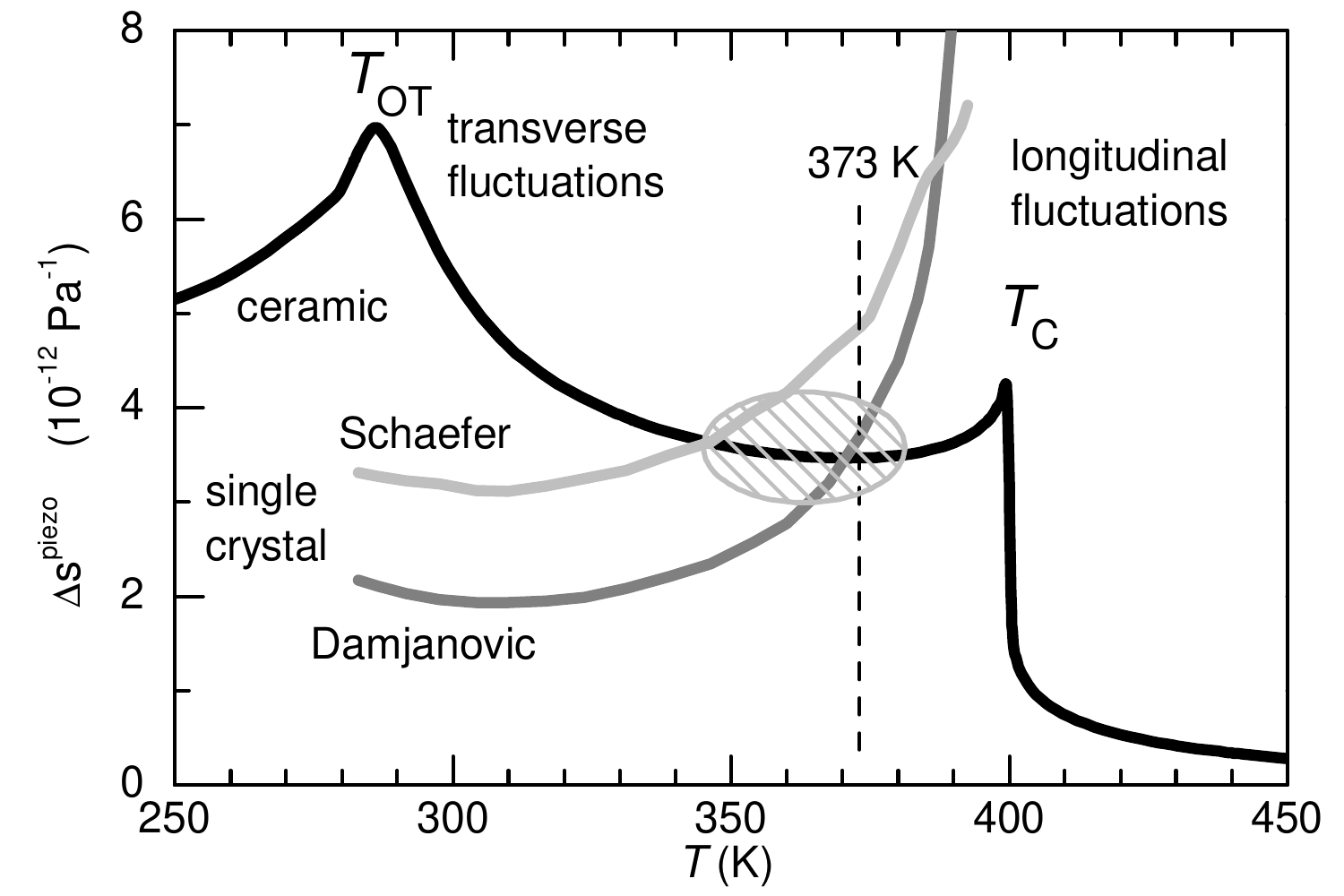}
\caption{Comparison between the piezoelectric softening of ceramic BT1,
after subtraction of the anharmonic background and rescaling from $%
s_{0}^{-1}=157$~GPa to $200$~GPa (see Fig. \protect\ref{fig_bkg}(b)), and
Eq. (\protect\ref{av-DspiezoT}) calculated with the single crystal data from
Damjanovic \textit{et al.} \protect\cite{DBS02,BDS03} and Schaefer \textit{%
et al. }\protect\cite{SSD86}}
\label{fig_comp}
\end{figure}

Encouraged by this result, we can test the validity of the method
outside the safe temperature range, around 373~K for BaTiO$_{3}$. The result
is shown in Fig. \ref{fig_comp}.

The black curve is the same $\Delta \mathbf{s}_{\text{ceram}}^{\text{piezo}}$
curve from ceramic BT1 as in Fig. \ref{fig_bkg}(b), and is compared with
those calculated from Eq. (\ref{av-DspiezoT}) with the $s_{\alpha \beta }$
and $\epsilon _{ij}$ constants from Damjanovic \textit{et al.}\cite%
{DBS02,BDS03} and Schaefer \textit{et al.}\cite{SSD86}. Even though both the
reference curves cross $\Delta \mathbf{s}_{\text{ceram}}^{\text{piezo}}$ in
the region of its plateau, hatched in Fig. \ref{fig_comp}, the agreement is
rather poor, because the reference data, especially those from the Landau
expansion, exhibit too large a rise on approaching $T_{\mathrm{C}}$ and too
small on approaching $T_{\mathrm{OT}}$. It seems unlikely that such large
differences can be totally attributed to the inadequacy of Eq. (\ref{Dspiezo}%
) in dealing with fluctuations, and there can be some weakness also in the
reference data.

For example, the $d_{i\beta }$ and $\epsilon _{ij}$ obtained from the Landau
potential \cite{Bel01} do not fully reproduce the experimental ones,
especially close to $T_{\mathrm{C}}$.\cite{WTD07} Those works were more
concerned with the anisotropy of the piezoelectric properties, rather than
on their temperature dependence, and were based on the Landau free energy
expansion of Bell\cite{Bel01} up to the sixth power of $P$. Wang \textit{et
al.}\cite{WTD07} have later shown that a more accurate representation of the
dielectric and ferroelectric properties of BaTiO$_{3}$ requires an expansion
up to the 8th order, with additional dependencies on temperature of some
coefficients. In particular, in this manner the temperature dependence of $%
\epsilon _{33}$ becomes steeper below $T_{\mathrm{C}}$ (Fig. 6 of Ref. \cite%
{WTD07}), mitigating the discrepancy between the calculated $\Delta \mathbf{s%
}_{\text{Landau}}^{\text{piezo}}$and $\Delta \mathbf{s}_{\text{ceram}}^{%
\text{piezo}}$. Another factor that worsens the comparison between these
curves\ below $T_{\mathrm{C}}$\ is the fact that the theoretical curves for $%
\Delta \mathbf{s}_{\text{Landau}}^{\text{piezo}}$were calculated with $T_{%
\mathrm{C}}=393$~K,\cite{DBS02,BDS03} while our sample has $T_{\mathrm{C}%
}=400$~K.

\section*{Conclusions}

The aim of this work was to verify that the softening of an unpoled ceramic
in the ferroelectric state with respect to the compliance of the
paraelectric state provides a measure of the piezoelectric coefficients,
based on the orientational average of Eq. (\ref{Dspiezo}) expressed in Eqs. (%
\ref{av-DspiezoT}-\ref{av-DspiezoO}). In order to obtain a quantitative
estimate, the other sources of softening, intrinsic and extrinsic, must be
considered, which are: \textit{i)} the association of the ferroelectric mode
with other modes, such as octahedral tilting in perovskites; \textit{ii)}
porosity; \textit{iii)} fluctuations; \textit{iv)} the anharmonic
stiffening of the elastic constants, linear in $T$.

If the ferroelectric transition is accompanied by another source of
spontaneous strain, it is probably impossible to disentangle the additional
softening from the piezoelectric one; the FE/AFE transition in PZT with
concomitant octahedral tilting is cited as a case where the two softenings
may be distinguished thanks to the different kinetics of the two modes.
Porosity appears as a major source of error; in the BaTiO$_{3}$ ceramic
samples considered here, the softening introduced by porosity can be
satisfactorily accounted for by rescaling the whole compliance curve by a
constant factor, but this simple correction may be inadequate for high
porosities with particular morphologies.

Barium titanate has been chosen for the experimental test, because its
intrinsic elastic and dielectric coefficients are available in the
literature, and because it undergoes a pure ferroelectric transition,
without the intervention of additional degrees of freedom; on the other
hand, BaTiO$_{3}$ exhibits important fluctuation effects and other FE
transitions, which complicate the analysis.

In the temperature region not too close to $T_{\mathrm{C}}$ and $T_{\mathrm{%
OT}}$, where fluctuations are expected to be negligible, there is good
agreement between the piezoelectric softening measured on ceramic samples
and the orientational average of the single crystal data, validating the
effectiveness of the elastic method for determining the piezoelectric
response, but the agreement worsens away from this region. Possible reasons
for the discrepancies in whole temperature range are discussed.

If the \textit{caveats} discussed above are observed, the evaluation of the
piezoelectric response from purely elastic measurements on unpoled ceramic
samples presents the great advantage of avoiding the procedure of poling,
and therefore of being insensitive to the uncertainties from partial poling
or thermal depoling, especially at high temperature. The method is
particularly promising in conjunction with the Resonant Ultrasound
Spectroscopy technique,\cite{TC15,ZTT17} since it allows all the elastic and
piezoelectric constants to be extracted with a single measurement through
the ferroelectric transition.

\bibliography{refs}



\end{document}